\documentclass{article}
\usepackage{amssymb}
\usepackage{amsmath}
\usepackage[english]{babel}
\usepackage{theorem}
\theorembodyfont{\upshape}

\newtheorem{theorem}{Theorem}

\newtheorem{corollary}[theorem]{Corollary}

\newtheorem{lemma}[theorem]{Lemma}

\newtheorem{proposition}[theorem]{Proposition}
\newtheorem{remark}[theorem]{Remark}
\begin{document}
\title{Spectral fluctuations of Schr\"odinger operators generated by critical points of the potential.}
\author{Brice Camus\\Ruhr-Universit\"at Bochum, Fakult\"at f\"ur
Mathematik,\\ Universit\"atsstr. 150, D-44780 Bochum,
Germany.\\
Email : brice.camus@univ-reims.fr} \maketitle
\begin{abstract}
\noindent  Starting from the spectrum of Schr\"odinger operators
on $\mathbb{R}^n$, we propose a method to detect critical points
of the potential. We argue semi-classically on the basis of a
mathematically rigorous version of Gutzwiller's trace formula
which expresses spectral statistics in term of classical orbits. A
critical point of the potential with zero momentum is an
equilibrium of the flow and generates certain singularities in the
spectrum. Via sharp spectral estimates, this fluctuation indicates
the presence of a critical point and allows to reconstruct
partially the local shape of the potential. Some
generalizations of this approach are also proposed.\medskip\\
keywords : Semi-classical analysis; Schr\"odinger operators;
Equilibriums in classical mechanics.
\end{abstract}
\section{Introduction.}
\textbf{Background.}

Let $P_{\hbar}=-\hbar^2\Delta+V$ be a Schr\"odinger operator where
the potential $V$ is smooth on $\mathbb{R}^n$ and bounded from
below. By a standard result in spectral theory, $P_{\hbar}$ has a
unique self-adjoint realization on a dense subset of
$L^2(\mathbb{R}^n)$. As usually, to this quantum operator
$P_{\hbar}$ we can associate a classical counterpart with the
Hamiltonian function $p(x,\xi)=\xi^2+V(x)$ on the phase space
$\mathbb{R}^n\times\mathbb{R}^n$. In what follows, we note
$\Phi_{t}$ the Hamiltonian flow of $H_{p}=\partial _{\xi
}p.\partial_{x} -\partial _{x}p.\partial_{\xi}$.

In the present contribution we are particularly interested in a
relation between the asymptotic properties of eigenvalues
$\lambda_j(\hbar)$ of $P_\hbar$ :
\begin{equation*}
P_{\hbar}\Psi_j(x,\hbar)=\lambda_j(\hbar) \Psi_j(x,\hbar), \text{
}\hbar\rightarrow 0,
\end{equation*}
and the closed orbits of $\Phi_{t}$. In geometry spectrum and
periodic orbits can be related, in a very explicit way, by means
of the Selberg and Duistermaat-Guillemin \cite{D-G} trace
formulae. In quantum mechanics, the existence of such a relation
is strongly suggested by the correspondence principle which
asserts that, in the semiclassical regime $\hbar \rightarrow 0$,
certain properties of $P_{\hbar}$ can be related to $\Phi_t$. In
the physic literature, a more precise formulation of this
principle appeared in the works of Balian\&Bloch \cite{BB} and
Gutzwiller \cite{GUT}. The Gutzwiller formula is usually stated
for the trace of the resolvent :
\begin{equation} \label{Gutzwiller}
\sum\limits_{j\in\mathbb{N}} \frac{1}{\lambda_j(\hbar)-E}\sim
\frac{\mathrm{Vol}(\Sigma_E)} {(2\pi \hbar)^n} +\frac{1}{i\hbar}
\sum\limits_{\gamma \in \Sigma_E} A_\gamma
e^{\frac{i}{\hbar}S_\gamma},
\end{equation}
where in the r.h.s the sum concerns the closed orbits $\gamma$
inside the surface
\begin{equation*}
\Sigma_E=\{ (x,\xi)\in \mathbb{R}^n\times\mathbb{R}^n \text{ / }
\xi^2+V(x)=E\}.
\end{equation*}
Also $\mathrm{Vol}(\Sigma_E)$ is the Riemannian volume of
$\Sigma_E$, $S_\gamma$ and $A_\gamma$ resp. the classical action
and the stability factor, including the Maslov phase, of $\gamma$.

In mathematics and in physics, such a relation between spectrum
and periodic orbits provides a powerful tool of analysis and
computation. See e.g. \cite{Laz} concerning the asymptotic
behavior of eigenvectors $\Psi_j(x,\hbar)$ and \cite{Haa} for
various applications in quantum chaos.
But, for a Schr\"odinger operator on $\mathbb{R}^n$, two different divergences occur in
Eq.(\ref{Gutzwiller}) :\medskip\\
1) The sum over the spectrum can be divergent. If the sum is
convergent it can also have a divergent behavior when
$\hbar\rightarrow 0$.\\
2) The sum over closed orbits is generally divergent. This is the
case if $|A_\gamma|$ does not decrease fast enough or if
the number of periodic orbits of period smaller than $T$ is exponentially growing with $T$.\medskip\\
\textbf{Mathematical approach of the Gutzwiller formula.}\\
As seen above, the question to remove divergences is important and
we explain below how to proceed. Assume that the spectrum of $P_h$
is discrete in some interval $[E-\varepsilon ,E+\varepsilon ]$, a
more global sufficient condition for this is given in section 2.
An accessible problem is to study the asymptotic behavior of the
spectral distributions :
\begin{equation}
\gamma (E,\hbar,\varphi )=\sum\limits_{|\lambda _{j}(\hbar)-E|\leq
\varepsilon }\varphi (\frac{\lambda _{j}(\hbar)-E}{\hbar}), \text{
as } \hbar\rightarrow 0, \label{Def trace}
\end{equation}
where $\varphi$ is a function chosen to remove the divergences. To
justify the terminology, observe that the truncated spectral
distribution :
\begin{equation*}
\sigma_{E,\varepsilon}(x)=\sum\limits_{|\lambda_j(\hbar)-E|\leq
\varepsilon} \delta_{\lambda_j(\hbar)}(x),
\end{equation*}
acting on a function $\varphi$ shifted by $E$ and scaled w.r.t.
$\hbar$ provides :
\begin{equation*}
\gamma(E,\hbar,\varphi)=\left\langle \sigma_{E,\varepsilon}(x) ,
\varphi(\frac{x-E}{\hbar}) \right \rangle.
\end{equation*}

In general, it is not possible to compute the spectrum of
$P_\hbar$ and one motivation is to derive statistics about
eigenvalues. For example, in Eq.(\ref{Def trace}) the formal
choice of $\varphi$ as the characteristic function of
$[-\eta,\eta]$, $0<\eta<\varepsilon$, determines the number
$N(\hbar)$ of bound states in $[E-\eta \hbar,E+\eta \hbar]$. Under
certain conditions, it can be proven that $N(\hbar)$ is
proportional to $\hbar^{1-n}$ (Weyl-law). Accordingly, for $n>1$
this implies that the finite sum defining
$\gamma(E,\hbar,\varphi)$ involves a large number of eigenvalues
as $\hbar\rightarrow 0$.

A second aspect is that the asymptotic expansion of
$\gamma(E,\hbar,\varphi)$ involves the classical dynamics in a
very explicit way. We recall that $E$ is regular if $\nabla
p(x,\xi )\neq 0$ on $\Sigma_{E}$ and critical otherwise, a
critical point $(x_0,\xi_0)$ of $p$ is a fixed point of $\Phi_t$
since $H_p(x_0,\xi_0)=0$. When $E$ is not critical the asymptotics
of Eq.(\ref{Def trace}) is well determined by the closed orbits of
$\Phi_t$ on $\Sigma_{E}$. For the full treatment of this problem,
and a complete formulation of the asymptotic expansion, we refer
to \cite{BU,PU}.

We explain now why the problem stated in Eq.(\ref{Def trace})
leads to a mathematically rigorous version of the Gutzwiller
formula. First, for each $\hbar>0$ the sum is finite and a
fortiori convergent. A convenient choice of $\varphi$ also ensures
that this quantity has an asymptotic expansion when
$\hbar\rightarrow 0$ independently from the choice of
$\varepsilon$. On the other side, it will be proven that only the
periods of $\Phi_t$ inside $\rm{supp}(\hat{\varphi})$, the support
of the Fourier transform of $\varphi$ :
\begin{equation}
\hat{\varphi}(t)=\int\limits_{\mathbb{R}} e^{itx}\varphi(x)dx,
\end{equation}
contribute in the asymptotic expansion. This principle is useful
since when $\rm{supp}(\hat{\varphi})$ is compact then finitely
many closed orbits of $\Sigma_E$ contribute and the second
divergence is solved. Hence if $\hat{\varphi}\in
C_0^{\infty}(\mathbb{R})$, the space of smooth functions with
compact support, $\varphi$ is in the Schwartz space
$\mathcal{S}(\mathbb{R})$. Since elements of
$\mathcal{S}(\mathbb{R})$ are smooth with exponential decay at
infinity, no divergence occurs and the size of $\varepsilon$ is
irrelevant in the semi-classical approximation.

Finally, in Eq.(\ref{Def trace}) the scaling w.r.t. $\hbar$ is
important. With this choice and via Fourier transform
considerations, we can use the propagator
$U_{\hbar}(t)=\exp(itP_{\hbar} /\hbar)$, solution of the
Schr\"odinger equation :
\begin{equation*}
-i\hbar \partial_t U_{\hbar}(t)=P_h U_{\hbar}(t),
\end{equation*}
to obtain a precise control w.r.t.$\hbar$. Roughly, $U_{\hbar}(t)$
can be expanded w.r.t. $\hbar$ via a so-called WKB approximation.
This expansion also provides the explicit relation with the
classical dynamics. The
precise technical justifications are given in section 3.\medskip\\
\textbf{Critical values and contributions of equilibrium.}

We have outlined the heuristic relation :
\begin{equation*}
\lim_{\hbar\rightarrow 0}\gamma (E,\hbar,\varphi
)\rightleftharpoons \{(t,x,\xi)\in\mathbb{R}\times \Sigma _{E}
\text{ / } \Phi_{t}(x,\xi)=(x,\xi)\}.
\end{equation*}
In the r.h.s any point $(x,\xi)$ of a periodic orbit appears only
at times $kT$, $k\in\mathbb{Z}$, where $T$ is the primitive period
of the orbit. But an equilibrium point $(x_0,\xi_0)$ satisfies
$\Phi_t(x_0,\xi_0)=(x_0,\xi_0)$ for all $t$. Hence when $E$ is no
more a regular value the nature of the set of fixed point changes
and some new contributions appear in the asymptotic expansion.

When $E=E_c$ is critical, the asymptotic behavior of Eq.(\ref{Def
trace}) is more complicated and is closely related to the geometry
of the flow inside $\Sigma_{E_c}$. For a non-degenerate critical
point, i.e. $d^{2}p$ is an invertible matrix when $dp=0$, the
reader can consult \cite{BPU}. The problem is treated there for
quite general operators, also including the case of a manifold of
critical points, but for $\rm{supp}(\hat{\varphi})$ small around
the origin. For Schr\"{o}dinger operators on $\mathbb{R}^n$ and
$\mathrm{supp}(\hat{\varphi})$ compact but arbitrary, the results
of \cite{BPU} are improved in \cite{KhD1}.

Two important problems occur in presence of critical points.
First, at every point where $dp=0$ the surface $\Sigma_{E_c}$ and
the metric of $\Sigma_{E_c}$ are not smooth. Next, the
determination of the asymptotic expansion w.r.t. $\hbar$ can be
very difficult. The point is that $\gamma(E,\hbar,\varphi)$ can be
expressed in terms of oscillatory integrals :
\begin{equation*}
I(\hbar)= \int\limits_{\mathbb{R}\times \mathbb{R}^{2n}}
a(t,x,\xi) e^{\frac{i}{\hbar} f(t,x,\xi)}dtdxd\xi,\text{ } \hbar
\rightarrow 0.
\end{equation*}
Note this oscillating factor $\hbar^{-1}$, precisely imposed by
the scaling w.r.t. $\hbar$ in Eq.(\ref{Def trace}). Via the WKB
approximation, the phase $f$ is related to the flow so that the
asymptotic behavior of $I(\hbar)$ is determined by the closed
orbits. The technical problem is that, in presence of an
equilibrium, $f$ has some degenerate critical points. The
stationary phase method cannot be applied and the asymptotic
expansion of $I(\hbar)$ is radically different : e.g. some terms
$\hbar^\alpha$, $\alpha\in\mathbb{Q}$ and powers of $\log(\hbar)$
generally appear in this setting.
\medskip\\
\textbf{Results and strategy}.

Our objective is to relate some variations in the discrete
spectrum of $P_{\hbar}$ with the presence of fixed points for the
classical system. Conversely, an attempt is made to prove that the
knowledge of such a spectral fluctuation can describe the
singularity of the potential. In theory, such a determination is
possible since the contributions of equilibriums are highly
sensitive to the local shape of $V$.

We will consider the case of a potential $V$ with finitely many
critical points $x_0^j$ attached to local homogeneous extremum of
$V$. An immediate consequence is that $p$ admits, locally, a
unique critical point $(x_0^j,0)$ on the surface
$\Sigma_{E_{c}^j}=\{(x,\xi)\in \mathbb{R}^{2n}\text{ / }
\xi^2+V(x)=V(x_0^j)\}$. A typical example is a polynomial double
well in dimension 1 where 3 critical points occur at the 2 minima
and at the maximum of V.

In fact, starting from a more precise relation :
\begin{equation*}
\lim_{\hbar\rightarrow 0}\gamma (E,\hbar,\varphi
)\rightleftharpoons
\{(t,x,\xi)\in\mathrm{supp}(\hat{\varphi})\times \Sigma _{E}
\text{ / } \Phi_{t}(x,\xi)=(x,\xi)\},
\end{equation*}
the core of the proof lies in 2 facts :\\
- Equilibriums have a continuous contribution w.r.t. the time
$t$.\\
- A convenient choice of $\mathrm{supp}(\hat{\varphi})$ erases all
other contributions.\medskip\\
Here, 'continuous contribution w.r.t. $t$' means that a fixed
point contribute to the asymptotic expansion of $\gamma
(E^j_c,\hbar,\varphi )$ in the form $\hbar^\alpha
\log(\hbar)^\beta \left\langle D,\hat{\varphi}\right\rangle$ with
$\mathrm{supp}(D)=\mathbb{R}$. Contrary to standard periodic
orbits whom contributions are supported in the set of periods,
such a term supported on the line cannot be erased just by
shrinking the support of $\hat{\varphi}$. For example, if
$\mathrm{supp}(\hat{\varphi})$ contains no period
of the flow the analysis easily follows if we view $\gamma(E,\hbar,\varphi)$ as a function of $E$ :\medskip\\
- The order w.r.t $\hbar$ of $\gamma(E,\hbar,\varphi)$ changes
when $E\rightarrow E^j_c$. This indicates the presence of an
equilibrium for $\Phi_t$, a fortiori of a critical point for $V$.\\
- This discontinuity at $E^j_c$ describes the shape of $V$.
\section{Hypotheses and main result.}
Let $p(x,\xi)=\xi^2 +V(x)$, where the potential $V$ is real valued
and smooth on $\mathbb{R}^n$. To this Hamiltonian is attached the
$\hbar$-differential operator $P_{\hbar}=-\hbar^2\Delta+V(x)$ and
by a classical result $P_{\hbar}$ is essentially self-adjoint if
$V$ is bounded from below.
\begin{remark}
\rm{We are here mainly interested in the case of Schr\"odinger
operators but a generalization to an $\hbar$-admissible operator
(e.g. in the sense of \cite{[Rob]}) of principal symbol
$\xi^2+V(x)$ is given in the last section.}
\end{remark}
First, to obtain a well defined spectral problem, we use
:\medskip\\
$(H_{1})$ $V\in C^{\infty}(\mathbb{R}^n)$. \textit{There exists } $C\in \mathbb{R}$ \textit{ such that }%
$\liminf\limits_{\infty}V >C$.\medskip\\
Note that $(H_1)$ is always satisfied if $V$ goes to infinity at
infinity. Now, consider an energy interval $J=[E_1,E_2]$ with
$E_2<\liminf\limits_{\infty}V$. In the following we note :
\begin{equation}
J(\varepsilon)=[E_1-\varepsilon,E_2+\varepsilon].
\end{equation}
For $\varepsilon<\varepsilon_0$ the set $p^{-1}(J(\varepsilon))$
is compact. By Theorem 3.13 of \cite{[Rob]} the spectrum $\sigma
(P_{\hbar})\cap J(\varepsilon)$ is discrete and consists in a
sequence $\lambda _{1}(\hbar)\leq \lambda _{2}(\hbar)\leq ...\leq
\lambda _{j}(\hbar)$ of eigenvalues of finite multiplicities, if
$\varepsilon$ and $\hbar$ are small enough.

The main tool of this work will be the spectral distribution :
\begin{equation}
\gamma (E,\hbar,\varphi)=\sum\limits_{\lambda _{j}(\hbar)\in
J(\varepsilon)}\varphi (\frac{\lambda _{j}(\hbar)-E}{\hbar}),
\label{Objet trace}
\end{equation}
more precisely, the asymptotic information contained in this
object. To avoid any problem of convergence we
impose the condition : \medskip\\
$(H_{2})$\textit{ We have }$\hat{\varphi}\in C_{0}^{\infty
}(\mathbb{R})$ \textit{ with a sufficiently small support near the
origin.}
\begin{remark}\rm{$(H_{2})$ is used
to erase contributions of non-trivial closed orbits and can be
relaxed to $\hat{\varphi}\in C_{0}^{\infty }(\mathbb{R})$ with a
weaker result. A more precise description of
$\mathrm{supp}(\hat{\varphi})$ is given in Lemma \ref{periods}.
For a non-degenerate minimum, it is more comfortable to assume
that $\mathrm{supp}(\hat{\varphi})$ contains no period of
$d\Phi_t(z_0)$. Some singularities, not relevant here, are
generated by these periods and we refer to \cite{BPU,KhD1} for a
detailed study of these contributions.}
\end{remark}
To simplify notations we write $z=(x,\xi)\in \mathbb{R}^{2n}$ and
$\Sigma_{E}=p^{-1}(\{E\})$  and we use the subscript $E_c$ to
distinguish out critical values of $p$. Of course one can also
work with
$T^*\mathbb{R}^n\simeq\mathbb{R}^{n}\times\mathbb{R}^{n}$. In $J$
there is finitely many critical values $E_c^1,...,E_c^l$ and in
$p^{-1}(J)$ finitely
many fixed points $z_0^1,...,z_0^m$, $m\geq l$. We impose now the type of singularity :\medskip\\
$(H_{3})$\textit{ On each }$\Sigma _{E_c^j}$ \textit{the symbol
}$p$\textit{ has isolated critical points }$z_{0}^j=(x_{0}^j,0).$
\textit{These critical points can be degenerate but are associated
to a local extremum of $V$ :
\begin{equation}\label{form pot}
V(x)=E_c+ V_{2k}(x)+\mathcal{O}(||x-x^j_0||^{2k+1}), \text{
}k\in\mathbb{N}^{*},
\end{equation}
where $V_{2k}$, homogeneous of degree $2k$, is definite positive
or negative.}
\begin{remark} \label{critical point}
\rm{For non-degenerate singularities we can apply the results of
\cite{BPU} and the extremum condition is not really necessary.}
\end{remark}
The next assumption, erases the mean values in the
trace formula :\medskip\\
$(H_4)$ \textit{$\hat{\varphi}$ is flat at 0, i.e. $\hat{\varphi}^{(j)}(0)=0$, $\forall j\in \mathbb{N}$}.\medskip\\
We could weaken $(H_4)$ to $\hat{\varphi}^{(j)}(0)=0$, $\forall
j\leq j_0$, where $j_0$ depends only on the degree of the
singularities of $V$ (cf.section 4). Note that such a $\varphi$
exists. Pick $g\in C_0^{\infty}(\mathbb{R})$,
$\mathrm{supp}(g)\subset [-M,M]$, then
$\hat{\varphi}(t)=t^{2j_0}g(t)$ satisfies our hypotheses. In this
case, we can pick $g$ even so that $\varphi$ is real.

Finally, to relax a bit $(H_2)$ we need a control on
the contribution of closed orbits. To do so, we impose the classical condition :\medskip\\
$(H_5)$ \textit{ All periodic trajectories of the flow are
non-degenerate.}\medskip\\
Non-degenerate closed orbits are those whose Poincar\'e map does
not admit 1 as eigenvalue and are isolated. The main result is :
\begin{theorem}\label{Main}
Assume $(H_{1})$ to $(H_{4})$ satisfied. As $\hbar$ tends to
$0^+$, we have :
\begin{equation*}
\gamma(s,\hbar,\varphi)= \left\{
\begin{matrix}
\mathcal{O}(\hbar^\infty) \text{ if } s\in [E_1,E_2]\backslash \{E_{c}^1,..., E_{c}^l\},\\
\mathcal{O}(f_j(\hbar)) \text{ if } s=E_{c}^j,\text { }j\in
\{1,...,l\},
\end{matrix}\right.
\end{equation*}
where each $f_j(\hbar)$ has a finite order w.r.t. $\hbar$.
\end{theorem}
Precisely, if $s=E_{c}^j$ carries a single minimum of degree $2k$
we obtain :
\begin{equation}
f_j(\hbar)=C(n,k,\varphi) \hbar^{\frac{n}{2}+\frac{n}{2k}-n}.
\end{equation}
But for a local maximum of $V$ we can obtain a logarithm of $h$ :
\begin{equation}
f_j(\hbar)=C(n,k,\varphi)
\hbar^{\frac{n}{2}+\frac{n}{2k}-n}\log(\hbar)^{j},\text{ } j=0
\text{ or } 1.
\end{equation}
In fact if the critical surface carries more than one critical
point then $f_j$ is the sum of their respective contributions.
Note that for $n=1$ and $k>1$ the singular term has negative order
w.r.t. $\hbar$. A more detailed formulation of each $f_j(\hbar)$
is given in Propositions \ref{minimum},\ref{maximum}. An
interesting property is that the singularity of
$\gamma(s,\hbar,\varphi)$ describes partially the singularity of
$V$.
\begin{corollary}\label{converse} Assume that $\Sigma_{E_c}$ carries exactly
one singularity $(x_0,0)$. Then the discontinuity of
$\gamma(s,\hbar,\varphi)$ at $s=E_c$ determines the degree of the
critical point and the spherical average of the germ of $V$ in
$x_0$.
\end{corollary}
This principle is limited in presence of multiple equilibriums on
the same surface since the sum of contributions of each critical
point could lead to a compensation. In $(H_2)$ the condition that
$\mathrm{supp}(\hat{\varphi})$ is small implies a very accurate
spectral estimate (e.g. by a Paley-Wiener estimates for
$\varphi$). It is possible to relax this assumption but the result
is weaker :
\begin{corollary}\label{weak} Assume $(H_1)$, $(H_3)$, $(H_4)$ and $(H_5)$
satisfied and that $\hat{\varphi}\in C_0^{\infty}(\mathbb{R})$,
then we obtain :
\begin{equation*}
\gamma(s,\hbar,\varphi)= \mathcal{O}(1) \text{ if } s\in
[E_1,E_2]\backslash \{E_c^1,..., E_c^l \}.
\end{equation*}
For critical values of $p$, estimates are the same as in Theorem
\ref{Main}.
\end{corollary}
The justification, given in section 4, is that in this case the
asymptotics is given by a finite sum over periodic orbits of
energy $s$. This result is weak if the singularity of $V$ is
non-degenerate since the equilibrium has a contribution of degree
0 w.r.t. $\hbar$. (cf. Propositions \ref{minimum},\ref{maximum} or
section 3 of \cite{BPU}). Finally, we would like to emphasize that
a maximum is more difficult to detect contrary to a local minimum
which is isolated on the energy surface.
\section{Oscillatory representation.}
The construction below is more or less classical and will be
sketchy. The only change is the choice of a more global
localization around $J=[E_1,E_2]$. Strictly speaking, with
$(H_1)$, we could also consider $]-\infty, E_2]$. Let be $\varphi
\in \mathcal{S}(\mathbb{R})$ with $\hat{\varphi}\in C_{0}^{\infty
}(\mathbb{R})$, we recall that :
\begin{equation*} \gamma (E,\hbar,\varphi)
=\sum\limits_{\lambda _{j}(\hbar)\in J(\varepsilon)}\varphi
(\frac{\lambda _{j}(\hbar)-E}{\hbar}), \text { }
J(\varepsilon)=[E_1-\varepsilon,E_2+\varepsilon],
\end{equation*}
with $p^{-1}(J(\varepsilon))$ compact in $T^{\ast
}\mathbb{R}^{n}$. For $\varepsilon>0$ small enough, we localize
around $J$ with a cut-off $\Theta \in C_{0}^{\infty
}(]E_1-\varepsilon ,E_2+\varepsilon \lbrack )$, such that $\Theta
=1$ on $J$ and $0\leq \Theta \leq 1$ on $\mathbb{R}$. We
accordingly split-up our spectral distribution as :
\begin{equation*}
\gamma (E,\hbar,\varphi)=\gamma _{1}(E,\hbar,\varphi)+\gamma
_{2}(E,\hbar,\varphi),
\end{equation*}
with :
\begin{gather*}
\gamma _{1}(E,\hbar,\varphi)=\sum\limits_{\lambda _{j}(\hbar)\in
J(\varepsilon)}(1-\Theta )(\lambda _{j}(\hbar))\varphi
(\frac{\lambda _{j}(\hbar)-E}{\hbar}),\\
\gamma _{2}(E,\hbar,\varphi)=\sum\limits_{\lambda _{j}(\hbar)\in
J(\varepsilon)}\Theta (\lambda _{j}(\hbar))\varphi (\frac{\lambda
_{j}(\hbar)-E}{\hbar}).
\end{gather*}
Since $\varphi\in \mathcal{S}(\mathbb{R})$ a classical estimate,
see e.g. Lemma 1 of \cite{Cam1}, is :
\begin{equation}
 \gamma _{1}(E,\hbar,\varphi)=\mathcal{O}(\hbar^{\infty }),
\text{ as } \hbar\rightarrow 0^{+}.\label{S1(h)=Tr}
\end{equation}
By inversion of the Fourier transform we have :
\begin{equation*}
\Theta (P_{\hbar})\varphi (\frac{P_{\hbar}-E}{\hbar})=\frac{1}{2\pi}\int\limits_{%
\mathbb{R}}e^{i\frac{tE}{\hbar}}\hat{\varphi}(t)\mathrm{exp}(-\frac{it}{\hbar}%
P_{\hbar})\Theta (P_{\hbar})dt.
\end{equation*}
The trace of the left hand-side is $\gamma _{2}(E,\hbar,\varphi)$
and Eq.(\ref{S1(h)=Tr}) provides :
\begin{equation}\label{Trace S2(h)}
\gamma (E,\hbar,\varphi)=\frac{1}{2\pi }\mathrm{Tr}\int\limits_{\mathbb{R}}e^{i%
\frac{tE}{\hbar}}\hat{\varphi}(t)\mathrm{exp}(-\frac{it}{\hbar}P_{\hbar})\Theta
(P_{\hbar})dt+\mathcal{O}(\hbar^{\infty }).
\end{equation}
Eq.(\ref{Trace S2(h)}) is very close to the classical Poisson
summation formula on $\mathbb{S}^1$ since the r.h.s. is expressed
below in term of the classical dynamics and this asymptotic
relation justifies the terminology of 'trace formula'. Moreover,
this formulation shows that the scaling w.r.t. $\hbar$, imposed in
the definition of $\gamma(E,\hbar,\varphi)$, is the best one since
we will solve the semi-classical propagator homogeneously w.r.t.
$\hbar$.

Let $U_{\hbar}(t)=\mathrm{exp}(-\frac{it}{\hbar}P_{\hbar})$ be the
quantum propagator. We approximate $U_{\hbar}(t)\Theta
(P_{\hbar})$ by a Fourier integral operator (FIO) depending on
$\hbar$. Let $\Lambda$ be the Lagrangian manifold associated to
the flow of $p$ :
\begin{equation*}
\Lambda =\{(t,\tau ,x,\xi ,y,\eta )\in T^{\ast }\mathbb{R}\times T^{\ast }%
\mathbb{R}^{n}\times T^{\ast }\mathbb{R}^{n}:\tau =p(x,\xi
),\text{ }(x,\xi )=\Phi _{t}(y,\eta )\},
\end{equation*}
and $I(\mathbb{R}^{2n+1},\Lambda )$ the class of oscillatory
integrals based on $\mathbb{R}^{2n+1}$ and whose Lagrangian
manifold is $\Lambda$. The next result is a semi-classical version
of a well known result on the propagator, see e.g. Duistermaat
\cite{DUI1}.
\begin{theorem}
The operator $U_{\hbar}(t)\Theta (P_{\hbar})$ is an $\hbar$-FIO
associated to $\Lambda$. For each $N\in\mathbb{N}$ there exists
$U_{\Theta ,\hbar}^{(N)}(t)$ with integral kernel in H\"ormander's
class $I(\mathbb{R}^{2n+1},\Lambda )$ and $R_{\hbar}^{(N)}(t)$
bounded, with a $L^{2}$-norm uniformly bounded for $0<\hbar\leq 1$
and $t$ in a compact subset of $\mathbb{R}$, such that :
\begin{equation*}
U_{\hbar}(t)\Theta(P_{\hbar})=U_{\Theta,\hbar}^{(N)}(t)+\hbar^{N}R_{\hbar}^{(N)}(t).
\end{equation*}
\end{theorem}
This result provides the existence of an asymptotic expansion in
power of $\hbar$ with a remainder that can be controlled since
$\mathrm{supp}(\hat{\varphi})$ is a compact. After perhaps a
reduction of $\varepsilon$, this remainder $R_{\hbar}^{(N)}(t)$ is
estimated via :
\begin{corollary}
Let $\Theta _{1}\in C_{0}^{\infty }(\mathbb{R})$, with $\Theta
_{1}=1$ on $\mathrm{supp}(\Theta )$ and $\mathrm{supp}(\Theta
_{1})\subset ]E_1-2\varepsilon, E_2+2\varepsilon[$, then $\forall
N\in \mathbb{N}$ :
\begin{equation*}
\mathrm{Tr}(\Theta (P_{\hbar})\varphi (\frac{P_{\hbar}-E}{\hbar}))=\frac{1}{2\pi }%
\mathrm{Tr}\int\limits_{\mathbb{R}}\hat{\varphi}(t)e^{\frac{i}{\hbar}%
tE}U_{\Theta ,\hbar}^{(N)}(t)\Theta
_{1}(P_{\hbar})dt+\mathcal{O}(\hbar^{N-n}).
\end{equation*}
\end{corollary}
For a proof of this result, based on the cyclicity of the trace
and a priori estimates on the spectral projectors (see
\cite{[Rob]}), we refer to \cite{Cam1}. For the particular case of
a Schr\"odinger operator the BKW ansatz shows that the integral
kernel of $U_{\Theta,\hbar}^{(N)}(t)$ can be recursively
constructed as :
\begin{gather*}
K_{\hbar}^{(N)}(t,x,y)=\frac{1}{(2\pi \hbar)^n}
\int\limits_{\mathbb{R}^n} b_{\hbar}^{(N)}(t,x,y,\xi)
e^{\frac{i}{\hbar} (S(t,x,\xi)-\left\langle y,\xi
\right\rangle)} d\xi,\\
b_{\hbar}^{(N)}=b_0+\hbar b_1+...+\hbar^N b_N,
\end{gather*}
where $S$ satisfies the Hamilton-Jacobi equation :
\begin{equation*}
p(x,\partial _{x}S(t,x,\xi))+\partial _{t}S(t,x,\xi )=0,
\end{equation*}
with initial condition $S(0,x,\xi)=\left\langle x,\xi
\right\rangle$. In particular we obtain that :
\begin{equation*}
\{(t,\partial _{t}S(t,x,\eta ),x,\partial _{x}S(t,x,\eta
),\partial _{\eta }S(t,x,\eta ),-\eta )\}\subset \Lambda ,
\end{equation*}
and that the function $S$ is a generating function of the flow,
i.e. :
\begin{equation}
\Phi _{t}(\partial _{\eta }S(t,x,\eta ),\eta ) =(x,\partial
_{x}S(t,x,\eta )). \label{Gene}
\end{equation}
We insert this approximation in Eq.(\ref{Trace S2(h)}), we set
$x=y$ and we integrate w.r.t. $x$. Modulo an error
$\mathcal{O}(\hbar^{N-n})$, we obtain that
$\gamma(E,\hbar,\varphi)$ equals :
\begin{equation}
\frac{1}{(2\pi \hbar)^n}\int\limits_{\mathbb{R}\times
T^*\mathbb{R}^n}e^{\frac{i}{\hbar}(S(t,x,\xi )-\left\langle x,\xi
\right\rangle +tE)}a_{\hbar}^{(N)} (t,x,\xi
)\hat{\varphi}(t)dtdxd\xi, \label{gamma1 OIF}
\end{equation}
where $a_{\hbar}^{(N)}(t,x,\eta )=b_{\hbar}^{(N)}(t,x,x,\eta )$.
\begin{remark}
\rm{By Theorem 3.11 \& Remark 3.14 of \cite{[Rob]}, $\Theta
(P_{\hbar})$ is $\hbar$-admissible. Moreover, the symbol is
compactly supported in
$p^{-1}([E_1-\varepsilon,E_2+\varepsilon])$. This point allows to
consider only oscillatory integrals with compact support for the
evaluation of the spectral distributions.$\hfill{\square}$}
\end{remark}
\textbf{Microlocalization of the trace.}\\
If $\psi\in C_0^{\infty}(T^{*}\mathbb{R}^{n})$, we recall that
$\psi^w(x,\hbar D_x)$ is the linear operator obtained by
Weyl-quantization of $\psi$, i.e. :
\begin{equation*}
\psi ^{w}(x,\hbar D_{x})f(x)=\frac{1}{(2\pi \hbar)^n}
\int\limits_{\mathbb{R}^{2n}} e^{\frac{i}{\hbar} \left\langle
x-y,\xi\right\rangle} \psi(\frac{x+y}{2},\xi) f(y)dyd\xi.
\end{equation*}
Mainly, the contribution of an equilibrium $z_0\in \Sigma_{E_c}$
can be reached via :
\begin{equation}\label{trace local}
\gamma_{z_0}(E_c,\hbar ,\varphi)=\frac{1}{2\pi }\mathrm{Tr}\int\limits_{\mathbb{R}}e^{i%
\frac{tE_{c}}{\hbar}}\hat{\varphi}(t)\psi ^{w}(x,\hbar D_{x})\mathrm{exp}(-\frac{i}{\hbar}%
tP_{\hbar})\Theta (P_{\hbar})dt,
\end{equation}
where $\psi\in C_0^{\infty} (T^* \mathbb{R}^n)$ is equal to 1 near
$z_0$. This principle will also be useful to obtain a weak
generalization in presence of multiple equilibriums.

We recall some basics on symbolic calculus with FIO. H\"ormander's
class of distributions with Lagrangian manifold $\Lambda$ over
$\mathbb{R}^n$ is noted $I(\mathbb{R}^{n},\Lambda )$. If
$(x_{0},\xi _{0})\in \Lambda $ and $\varphi (x,\theta )\in
C^{\infty }(\mathbb{R}^{n}\times \mathbb{R}^{N})$ parameterizes
$\Lambda $ in a sufficiently small neighborhood $U$ of $(x_{0},\xi
_{0})$, then for each $u_{\hbar}\in I(\mathbb{R}^{n},\Lambda )$
and $\chi \in C_{0}^{\infty }(T^{\ast }\mathbb{R}^{n})$,
$\rm{supp}(\chi )\subset U,$ there exists a
sequence of amplitudes $c_{j}(x,\theta )\in C_{0}^{\infty }(\mathbb{R}%
^{n}\times \mathbb{R}^{N})$ such that for all $L\in\mathbb{N}$ :
\begin{equation*}
\chi ^{w}(x,\hbar D_{x})u_{\hbar}=\sum\limits_{-d\leq j<L}\hbar^{j}I(c_{j}e^{\frac{i}{\hbar}%
\varphi })+\mathcal{O}(\hbar^{L}).
\end{equation*}
Hence, for each $N\in \mathbb{N}^{*}$ and modulo an error
$\mathcal{O}(\hbar^{N-d})$, the localized trace
$\gamma_{z_0}(E_c,\hbar,\varphi)$ of Eq.(\ref{trace local}) can be
written as :
\begin{equation}
(2\pi \hbar)^{-d}\int\limits_{\mathbb{%
R\times R}^{2n}}e^{\frac{i}{\hbar}(S(t,x,\xi )-\left\langle x,\xi
\right\rangle +tE_c)}\tilde{a}_{\hbar}^{(N)}(t,x,\xi
)\hat{\varphi}(t)dtdxd\xi . \label{gamma1 OIF2}
\end{equation}
To get the right power $-d$ of $\hbar$, we apply results of
Duistermaat \cite{DUI1} on the order : $\hbar$-pseudo-differential
operators $\psi^w(x,\hbar D_x)$ are of order 0 w.r.t. $1/\hbar$.
Since the order of $U_{\hbar}(t)\Theta (P_{\hbar})$ is
$-\frac{1}{4}$, we have :
\begin{equation*}
\psi ^{w}(x,\hbar D_{x})U_{\hbar}(t)\Theta (P_{\hbar})\sim (2\pi
\hbar)^{-n}\int\limits_{\mathbb{R}^{n}}\tilde{a}_{\hbar}^{(N)}(t,x,y,\eta )e^{\frac{i}{\hbar}%
(S(t,x,\eta )-\left\langle y,\eta \right\rangle )}dy.
\end{equation*}
Multiplying by $\hat{\varphi}(t)e^{\frac{i}{h}tE_{c}}$ and passing
to the trace we find Eq.(\ref{gamma1 OIF2}) with $d=n$ and we
write again $\tilde{a}_{\hbar}^{(N)}(t,x,\eta )$ for
$\tilde{a}_{\hbar}^{(N)}(t,x,x,\eta )$. In particular :
\begin{equation}
\tilde{a}_{\hbar}^{(0)}(t,x,x,\eta )= \psi(x,\eta)
a_{0}(t,x,x,\eta),
\end{equation}
is independent of $\hbar$ and is compactly supported w.r.t.
$(x,\eta)$.
\section{Proof of the main result.}
\textbf{Classical dynamics near the equilibrium.}\\
A generic critical points of the phase function of Eq.(\ref{gamma1
OIF}) satisfies :
\begin{equation*}
\left\{
\begin{array}{c}
E=-\partial _{t}S(t,x,\xi ), \\
x=\partial _{\xi }S(t,x,\xi ), \\
\xi =\partial _{x}S(t,x,\xi ),
\end{array}
\right. \Leftrightarrow \left\{
\begin{array}{c}
p(x,\xi )=E, \\
\Phi _{t}(x,\xi )=(x,\xi ),
\end{array}
\right.
\end{equation*}
where the right hand side defines a closed trajectory of the flow
inside $\Sigma_{E}$. Note that equilibriums are also included. By
the non-stationary phase lemma, outside of the critical set the
contribution is $\mathcal{O}(\hbar^{\infty})$.

Let be $E_c$ any critical value in $[E_1,E_2]$ and $z_0$ an
equilibrium of $\Sigma_{E_c}$. We choose a function $\psi \in
C_{0}^{\infty }(T^{\ast }\mathbb{R}^{n})$, with $\psi =1\text{
near }z_{0}$, hence :
\begin{gather*}
\gamma _{2}(E_{c},\hbar,\varphi) =\frac{1}{2\pi }\mathrm{Tr}\int\limits_{\mathbb{R}}e^{i%
\frac{tE_{c}}{\hbar}}\hat{\varphi}(t)\psi ^{w}(x,\hbar D_{x})\mathrm{exp}(-\frac{i}{\hbar}%
tP_{\hbar})\Theta (P_{\hbar})dt\\
+\frac{1}{2\pi }\mathrm{Tr}\int\limits_{\mathbb{R}}e^{i\frac{tE_{c}}{\hbar}}\hat{%
\varphi}(t)(1-\psi ^{w}(x,\hbar
D_{x}))\mathrm{exp}(-\frac{i}{h}tP_{\hbar})\Theta (P_{\hbar})dt.
\end{gather*}
If there is no other singularity on $\Sigma_{E_c}$ with $(H_5)$
the asymptotic expansion of the second term is given by the
semi-classical trace formula on a regular level. For finitely many
critical point on $\Sigma_{E_c}$, we can repeat the procedure. The
first term is micro-local and precisely generate the singularity
in Theorem \ref{Main}. We note $\Omega$ the discrete set of
critical points $z_0^j$ in $p^{-1}(J)$. The next result provides a
global information on the periods of the classical flow.
\begin{lemma}\label{periods} There exists a $T>0$, depending only on $V$ and $J=[E_1,E_2]$,
such that $\Phi_t(z)\neq z$ for all $z\in p^{-1}(J)\backslash
\Omega$ and all $t\in]-T,0[\cup ]0,T[$.
\end{lemma}
\textit{Proof.} If $H_p$ is our hamiltonian vector field and
$z=(x,\xi)$ we have :
\begin{equation*}
|| H_p(z_1)-H_p(z_2)||^2= 4||\xi_1-\xi_2||^2
+||\nabla_xV(x_1)-\nabla_xV(x_2)||^2.
\end{equation*}
When $z_1$ and $z_2$ are in the compact $p^{-1}(J)$ there exists
$b>0$ such that :
\begin{equation*}
||\nabla_xV(x_1)-\nabla_xV(x_2)||\leq b ||x_1-x_2||.
\end{equation*}
Hence, there exists $a>0$ such that :
\begin{equation*}
||H_p(z_1)-H_p(z_2)||\leq  a ||z_1-z_2||,\text { }\forall z_1,z_2
\in p^{-1}(J).
\end{equation*}
The main result of \cite{Yor} shows that any periodic orbit inside
$p^{-1}(J)$ has a period $\tau \geq 2\pi/a>0$. The lemma follows
with $T:=T(V,J)=2\pi/a$. $\hfill{\blacksquare}$
\begin{remark} \rm{The result of \cite{Yor} is optimal (harmonic
oscillator). Note that $T$ is decreasing if one increase the size
of $J$. Lemma \ref{periods} provides a total control on the r.h.s.
of the trace formula : if $\hat{\varphi}\in C_0^{\infty}(]-T,T[)$,
the only contribution arises from the set $\{(t,z_0),\text{
}t\in\mathrm{supp}(\hat{\varphi})\}$.}
\end{remark}
Now, we restrict our attention to the singular contribution
generated by one critical point. As pointed out in section 2, for
a non degenerate extremum a minor technical problem could occur.
We recall that the linearized flow $d\Phi_t$ is the differential
of the flow $\Phi_t$ w.r.t. initial conditions $z=(x,\xi)$. When
$z_0$ is a critical point of $p$, the linear map $z\mapsto
d\Phi_t(z_0)z$ can be interpreted as the Hamiltonian flow of
$z\mapsto \langle d^2p(z_0)z,z\rangle$. After perhaps a change of
local coordinates near $x_0$, we can assume that $d^2V(x_0)$ is
diagonal. If $x_0$ is a maximum of the potential $d\Phi_t(z_0)$
has no non-zero period which ends immediately the discussion. If
$x_0$ is a minimum $d\Phi_t(z_0)$ is elliptic with primitive
periods $(T_1,..,T_n)$ generated by the eigenvalues of $d^2
V(x_0)$. But the constant $b$ of Lemma \ref{periods} is certainly
bigger than the spectral radius of $d^2 V(x_0)$ and hence we have
the inequality $T<\min \{ T_1,..,T_n\}$. Following the approach of
\cite{BPU} or \cite{KhD1}, if $\mathrm{supp}(\hat{\varphi})\subset
]-T,T[$ the associated contribution is smooth on
$\mathrm{supp}(\hat{\varphi})\backslash \{0\}$. For a degenerate
critical point as in $(H_3)$ a surprising result, established in
\cite{Cam3,Cam4}, is that the only singularity is located at
$t=0$. Hence no extra
assumption on $\hat{\varphi}$ is required.\medskip\\
\textbf{The trace as an energy function.}\\
As seen in section 2 it suffices to study the localized problem :
\begin{equation*}
\gamma _{z_{0}}(E_{c},\hbar,\varphi)=\frac{1}{2\pi }\mathrm{Tr}\int\limits_{\mathbb{R}}e^{i%
\frac{tE_{c}}{\hbar}}\hat{\varphi}(t)\psi ^{w}(x,\hbar D_{x})\mathrm{exp}(-\frac{it}{\hbar}%
P_{\hbar})\Theta (P_{\hbar})dt.
\end{equation*}
Here $\psi \in C_{0}^{\infty }(T^{\ast }\mathbb{R}^{n})$ is
micro-locally supported near $z_{0}$ (cf section 2). For the
convenience of the reader we recall the contributions of
equilibriums in the trace formula. We note
$\mathrm{S}(\mathbb{S}^{n-1})$ the surface of $\mathbb{S}^{n-1}$
and in the next two propositions it is understood that conditions
$(H_1)$ to $(H_3)$ are satisfied.
\begin{proposition}\label{minimum}
If $x_0$ is a local minimum we have :
\begin{equation*}
\gamma _{z_{0}}(E_{c},\hbar,\varphi)\sim
\hbar^{\frac{n}{2}+\frac{n}{2k}-n}
\sum\limits_{j,l\in\mathbb{N}^2} \hbar^{\frac{j}{2}+\frac{l}{2k}}
\Lambda_{j,l}(\varphi ),
\end{equation*}
where the $\Lambda_{j,l}$ are some distributions. The leading
coefficient is :
\begin{equation*}
\hbar^{\frac{n}{2}+\frac{n}{2k}-n}
\frac{\mathrm{S}(\mathbb{S}^{n-1})}{(2\pi)^n}
\int\limits_{\mathbb{S}^{n-1}} |V_{2k}(\eta)|^{-\frac{n}{2k}}
d\eta \int\limits_{\mathbb{R}_{+} \times \mathbb{R}_{+}}
\varphi(u^2 +v^{2k}) u^{n-1} v^{n-1} dudv.
\end{equation*}
\end{proposition}
\begin{proposition}\label{maximum}
If $x_0$ is a local maximum we have :
\begin{equation*}
\gamma _{z_{0}}(E_{c},\hbar,\varphi)\sim
\hbar^{\frac{n}{2}+\frac{n}{2k}-n}
\sum\limits_{m=0,1}\sum\limits_{j,l\in\mathbb{N}^2}
\hbar^{\frac{j}{2}+\frac{l}{2k}}\mathrm{log}(\hbar)^m \Lambda
_{j,l,m}(\varphi ).
\end{equation*}
If $\frac{n(k+1)}{2k}\notin \mathbb{N}$, the first non-zero
coefficient is given by :
\begin{equation*}
\hbar^{\frac{n}{2} +\frac{n}{2k}-n}%
\left\langle T_{n,k},\varphi\right\rangle%
\frac{\mathrm{S}(\mathbb{S}^{n-1})}{(2\pi)^n}
\int\limits_{\mathbb{S}^{n-1}}
|V_{2k}(\eta)|^{-\frac{n}{2k}}d\eta.
\end{equation*}
The distributions $T_{n,k}$ are respectively given by :
\begin{gather*}
\left\langle T_{n,k},\varphi\right\rangle
=\int\limits_{\mathbb{R}}
(C_{n,k}^{+}|t|_{+}^{n\frac{k+1}{2k}-1}+C_{n,k}^{-}|t|_{-}^{n\frac{k+1}{2k}-1})
\varphi (t) dt,
\text{ if }n \text{ is odd},\\
\left\langle T_{n,k},\varphi\right\rangle
=C_{n,k}^{-}\int\limits_{\mathbb{R}} |t|_{-}^{n\frac{k+1}{2k}-1}
\varphi (t) dt,\text{ if }n \text{ is even}.
\end{gather*}
But if $\frac{n(k+1)}{2k}\in \mathbb{N}$ and $n$ is odd then the
top-order term is :
\begin{equation*}
C_{n,k} \log (\hbar)\hbar^{\frac{n}{2}+\frac{n}{2k}-n}%
\frac{\mathrm{S}(\mathbb{S}^{n-1})}{(2\pi)^n}\int\limits_{\mathbb{S}^{n-1}}
|V_{2k}(\eta)|^{-\frac{n}{2k}} d\eta  \int\limits_{\mathbb{R}}
|t|^{n\frac{k+1}{2k}-1} \varphi (t) dt.
\end{equation*}
Finally, if $\frac{n(k+1)}{2k}\in \mathbb{N}$ and $n$ is even,
$C_{n,k}^{+}=C_{n,k}^{-}$ and we have :
\begin{equation*}
C_{n,k}^{\pm} \hbar^{\frac{n}{2}+\frac{n}{2k}-n}%
\frac{1}{(2\pi)^n} \int\limits_{\mathbb{S}^{n-1}}
|V_{2k}(\eta)|^{-\frac{n}{2k}}d\eta \int\limits_{\mathbb{R}}
|t|^{n\frac{k+1}{2k}-1} \varphi (t) dt.
\end{equation*}
\end{proposition}
\begin{remark} \rm{To emphasize the consistency of these results we precise that
$C_{n,k}$, $C_{n,k}^{\pm}$ are non-zero universal constants
depending only on $n$ and $k$. Such terms $\hbar^\alpha$ and
$\hbar^\alpha \log(\hbar)$, $\alpha\in \mathbb{Q}$ never appear if
$E$ is regular.}
\end{remark}
For a proof we refer to \cite{Cam3,Cam4} resp. for a local minimum
and maximum. The case $k=1$ was treated in \cite{BPU}. With
$(H_5)$ and $E$ regular, we have :
\begin{gather*}
\gamma(E,\hbar,\varphi)%
\sim \frac{\hbar^{1-n}}{(2\pi)^n} \mathrm{LVol}(\Sigma_E)
\hat{\varphi}(0)+\sum\limits_{j=1}^{\infty} \hbar^{1-n+j}
c_j(\hat{\varphi})(0)\\
+\sum\limits_{\rho\in \Sigma_E} e^{\frac{i}{\hbar}S_\rho} e^{i\pi
\mu_\rho/4} \sum\limits_{j=0}^{\infty}D_{\rho,j}
(\hat{\varphi})(T_\rho) \hbar^j.
\end{gather*}
We refer to \cite{PU} for a proof. In the r.h.s. the sum concerns
periodic orbits $\rho$ of energy $E$ and is finite since
$\mathrm{supp}(\hat{\varphi})$ is compact. Here $S_\rho$,
$\mu_\rho$ and $T_\rho$ are resp. the action, the Maslov-index and
the period of the closed orbit $\rho$ and both $c_j$, $D_{\rho,j}$
are differential operators of order $j$. If $\varphi$ satisfies
$(H_4)$ we have $c_j(\hat{\varphi})(0)=0$ and for each
$s\in[E_1,E_2]$ regular :
\begin{equation}\label{sum orbits}
\gamma(s,\hbar,\varphi)%
\sim \sum\limits_{\rho\in\Sigma_s} e^{\frac{i}{\hbar}S_\rho}
e^{i\pi \mu_\rho/4} \sum\limits_{j=0}^{\infty} D_{\rho,j}
(\hat{\varphi})(T_\rho) \hbar^j.
\end{equation}
We accordingly obtain that :
\begin{equation}\label{trace non critical}
\gamma(s,\hbar,\varphi)=\mathcal{O}(1),\text{ } \forall s\in
[E_1,E_2]\backslash\{E_c^1,...,E_c^l\}.
\end{equation}
This point will justify Corollary \ref{weak}. By Lemma
\ref{periods}, we have $T_\rho\geq T$ uniformly w.r.t.
$s\in[E_1,E_2]$. Hence if $s$ is not critical and $(H_2)$ is
satisfied the sum over the periods of Eq.(\ref{sum orbits}) is
simply 0 and in Eq.(\ref{trace non critical}) we obtain in fact
$\mathcal{O}(h^\infty)$. Note that $(H_5)$ is not required here.
For $s=E_c^m$ critical there is a continuous contribution w.r.t.
$t$ in the spectral distribution. A fortiori, a choice of
$\hat{\varphi}$ flat at the origin does not erase this term. We
have :
\begin{equation*}
\gamma(E_{c}^m,\hbar,\varphi)  \sim \sum\limits_{j=1}^{N_m}
f_j(\hbar),
\end{equation*}
where $N_m$ is the number of equilibrium on $\Sigma_{E_{c}^m}$ and
each $f_j(\hbar)$ is given by the leading term of
Propositions \ref{minimum},\ref{maximum}.$\hfill{\blacksquare}$\medskip\\
Note that the bottom of a symmetric double well gives a similar
answer as a single well of same nature. Hence without microlocal
considerations it is difficult to distinguish
these 2 different settings.\medskip\\
\textit{Proof of corollary \ref{converse}.} First, the Weyl-law
for regular energies :
\begin{equation*}
\gamma(E,\hbar,\varphi) \sim (2\pi \hbar)^{1-n} \hat{\varphi}(0)
\mathrm{Lvol}(\Sigma_E),
\end{equation*}
computes the dimension $n$. Now assume given a critical value
$E_c$ with a single critical point. The only choice of the
spectral function $\varphi$ allows to detect $E_c$ via the
singularity $f(\hbar)$ of Theorem \ref{Main}. The knowledge of
$f(\hbar)$ determines the order of the contribution. For example,
if :
\begin{equation*}
f(\hbar)\sim C \hbar^\alpha \log(\hbar),
\end{equation*}
the critical point is a maximum and $\alpha$ computes the degree
$2k$ of the singularity. With $\hat{\varphi}$, the knowledge of
$k$ allows to compute the quantity :
\begin{equation*}
\int\limits_{\mathbb{R}} |t|^{n\frac{k+1}{2k}-1} \varphi (t) dt.
\end{equation*}
A fortiori $C$ determines the average of
$|V_{2k}|^{-\frac{n}{2k}}$ on $\mathbb{S}^{n-1}$. Without
$\log(\hbar)$, the nature of the critical point can be detected by
a symmetry argument w.r.t. $\varphi$ since we a priori know $n$
and $k$. In view of Propositions \ref{minimum},\ref{maximum} we
can choose $\varphi$ odd, even, symmetric or non-symmetric w.r.t.
the origin to conclude. Note that if $\hat{\varphi}$ is not even
$\varphi$ is a priori complex valued. $\hfill{\blacksquare}$\medskip\\
The spherical average of $V_{2k}$ is a Jacobian. For example we
have :
\begin{equation*}
\int\limits_{\mathbb{R}^n} e^{-|V_{2k}(x)|}dx
=\frac{1}{2k}\Gamma(\frac{n}{2k}) \int\limits_{\mathbb{S}^{n-1}}
|V_{2k}(\eta)|^{-\frac{n}{2k}} d\eta.
\end{equation*}
A similar result holds for the pullback $f(V_{2k}(x))$, if $f\in
L^1(\mathbb{R}_{+},r^{\frac{n}{2k}-1}dr)$.
\begin{remark} \rm{Enlarging the list of
singularities would provide a bigger "dictionary". The case of
non-homogeneous singularities for $V$ is still an open problem, in
particular because the determination of an explicit asymptotic
expansion w.r.t. $\hbar$ can be very difficult.}
\end{remark}
We propose now 2 slight generalizations of the main result.\medskip\\
\textbf{a) Effect of a sub-principal symbol.}\\
Because of some recent developments of Helffer\&Sj\"ostrand for
Witten Laplacians, see e.g. \cite{Hel} for an overview and
references,  we show shortly how to extend the result of Theorem
\ref{Main} to the case of an $\hbar$-admissible operator. For
example, the Witten Laplacian on zero-forms is :
\begin{equation*}
\Delta_{f,\hbar}^{(0)}=-\hbar^2 \Delta+ \frac{1}{4} |\nabla
f(x)|^2 -\frac{\hbar}{2} \Delta f(x), \text{ } f\in
C^{\infty}(\mathbb{R}^n),
\end{equation*}
whose symbol $p(x,\xi)=p_0(x,\xi)+\hbar p_1(x,\xi)$ depends on
$\hbar$. More generally, it is possible to consider operators
$P_\hbar$ of symbol $p_\hbar\sim \sum \hbar^j p_j$ (Borel sum)
with principal symbol $p_0(x,\xi)=\xi^2+V(x)$ and a subprincipal
symbol $p_1\neq 0$. Starting from the results of section 3 we
proceed as follows.

To each element $u_{\hbar}$ of $I(\mathbb{R}^{n},\Lambda )$ we can
associate canonically a principal symbol
$e^{\frac{i}{\hbar}S}\sigma _{\mathrm{princ}}(u_{\hbar})$, where
$S$ is a function on $\Lambda $ such that $\xi dx=dS$ on
$\Lambda$. In fact, if $u_{\hbar}$ can locally be represented by
an oscillatory integral with amplitude $a$ and phase $\varphi$,
then we have $S=S_{\varphi }=\varphi \circ i_{\varphi }^{-1}$ and
$\sigma _{\mathrm{princ}}(u_{\hbar})$ is a section of $|\Lambda
|^{\frac{1}{2}} \otimes M(\Lambda )$, where $M(\Lambda )$ is the
Maslov vector-bundle of $\Lambda $ and $|\Lambda |^{\frac{1}{2}}$
the bundle of half-densities on $\Lambda$. When $p_{1}\neq 0$, in
the global coordinates $(t,y,\eta )$ on $\Lambda$, the
half-density of $U_{\hbar}(t)$ is given by :
\begin{equation}
\nu(t,y,\eta)=\exp (i\int\limits_{0}^{t}p_{1}(\Phi _{s}(y,-\eta ))ds)|dtdyd\eta |^{\frac{1%
}{2}}.\label{demi densite}
\end{equation}
For this expression, related to the resolution of the first
transport equation for the propagator, we refer to Duistermaat and
H\"{o}rmander \cite{D-H}. Accordingly, the FIO approximating the
propagator has the amplitude :
\begin{equation*}
\tilde{a}(t,z)= a(t,z)\exp (i \int\limits_{0}^{t}
p_1(\Phi_s(z))ds).
\end{equation*}
Since $z_0$ is an equilibrium we have $p_1(\Phi_s(z_0))=p_1(z_0)$,
$\forall s$, and :
\begin{equation}
\tilde{a}(t,z_0)= \hat{\varphi}(t) e^{it p_1(z_0)}.
\end{equation}
If the subprincipal symbol vanishes at the critical point, which
is the case in a lot of practical situations, the trace formula
remains the same. If $p_1(z_0)\neq 0$, by Fourier inversion
formula we replace $\varphi(t)$ by $\varphi(t+p_1(z_0))$ in all
integral formulae of Propositions \ref{minimum},\ref{maximum}.
Note that, with $(H_4)$,
this has absolutely no effect for the mean values and hence on the main result.\medskip\\
\textbf{b) A micro-local approach.}\\
We inspect now the case of an energy surface supporting more than
one critical point, but with a much more restrictive method. Let
be $K=p^{-1}(J)\subset T^{*}\mathbb{R}^n$ and
$r_0=\frac{1}{2}\inf\limits_{i\neq j} d(z_i,z_j)$, where $d$ is
any distance on $T^*\mathbb{R}^n$. Each open ball $B(z,r_0)\subset
T^{*}\mathbb{R}^n$ contains at most 1 critical point for each
$z\in K$. Clearly, we can cover a compact neighborhood of $K$ by a
finite number of balls $B(z,r_0)$. With a partition of unity,
adapted to this covering, we obtain :
\begin{equation*}
\sum\limits_{j=1}^{N} \psi_j^w(x,\hbar D_x) =\mathrm{Id}, \text{
on } C_0^{\infty}(K).
\end{equation*}
For each $s\in J$, we obtain :
\begin{equation*}
\mathrm{Tr}\int\limits_{\mathbb{R}} \hat{\varphi}(t)
\Theta(P_\hbar)e^{\frac{i}{\hbar}t(P_{\hbar}-s)}dt
=\sum\limits_{j=1}^{N} \mathrm{Tr}\int\limits_{\mathbb{R}}
\hat{\varphi}(t) \psi_j^w(x,\hbar
D_x)\Theta(P_\hbar)e^{\frac{i}{\hbar}t(P_{\hbar}-s)}dt.
\end{equation*}
Note that the r.h.s. is studied in section 2. By the same argument
as before, if $\Sigma_s\cap \mathrm{supp}(\psi_j)$ contains no
critical point we obtain :
\begin{equation*}
\mathrm{Tr}\int\limits_{\mathbb{R}} \hat{\varphi}(t)
\psi_j^w(x,\hbar D_x)\Theta(P_\hbar)e^{\frac{i}{\hbar}t(P_{\hbar}-s)}dt=%
\mathcal{O}(\hbar^{\infty}).\\
\end{equation*}
And if there is exactly one critical point $z_0\in\Sigma_s$ in
$\mathrm{supp}(\psi_j)$ we have :
\begin{equation*}
\mathrm{Tr}\int\limits_{\mathbb{R}} \hat{\varphi}(t)
\psi_j^w(x,\hbar
D_x)\Theta(P_{\hbar})e^{\frac{i}{\hbar}t(P_{\hbar}-s)}dt=
\psi(z_0){f}_j(\hbar),
\end{equation*}
and by construction no cancellation can occur.
\begin{remark}\rm{In Corollary \ref{weak} we have considered
$(H_5)$ for the flow. A similar result holds for a chaotic
dynamics and an isolated degenerate closed orbit can be treated as
in \cite{Pop}. Finally, using the results of \cite{PU} one can
extend Corollary \ref{weak} to the case of families of periodic
orbits of dimension $d\leq n$.}
\end{remark}
\textbf{Acknowledgments.} This work was supported by the
\textit{SFB-TR12}, \textit{Symmetries and Universality in
Mesoscopic Systems}.

\end{document}